\documentclass[sigplan,screen]{acmart}
\pdfoutput=1

\setcopyright{cc}
\setcctype{by}
\acmDOI{10.1145/3696443.3708920}
\acmYear{2025}
\copyrightyear{2025}
\acmISBN{979-8-4007-1275-3/25/03}
\acmConference[CGO '25]{Proceedings of the 23rd ACM/IEEE International Symposium on Code Generation and Optimization}{March 01--05, 2025}{Las Vegas, NV, USA}
\acmBooktitle{Proceedings of the 23rd ACM/IEEE International Symposium on Code Generation and Optimization (CGO '25), March 01--05, 2025, Las Vegas, NV, USA}
\received{2024-05-23}
\received[accepted]{2024-07-22}

\AtBeginDocument{
  }

\AtBeginDocument{%
  }

\usepackage{amsmath}
\usepackage{algorithmic}
\usepackage{graphicx}
\usepackage{textcomp}
\usepackage{xcolor}
\usepackage{comment}

\usepackage[inline]{enumitem}
\usepackage{url}
\usepackage{balance}
\newcommand{\myfontsizealgorithms}{\fontsize{7}{7}\selectfont}
\usepackage[ruled,vlined,linesnumbered]{algorithm2e}
\usepackage{grumble}
\usepackage{natbib}

\newcommand{\myparagraph}[1]{\noindent{\bf {#1}.}}

\newcommand{\out}[1] {}

\newcounter{codeLineCntr}

\newcommand{\punt}[1]{}

\renewcommand{\eqref}[1]{Equation~(\ref{eq:#1})}

\newcounter{remark}[section]

\definecolor{mygreen}{rgb}{0,0.6,0}
\definecolor{mygray}{rgb}{0.5,0.5,0.5}
\definecolor{mymauve}{rgb}{0.58,0,0.82}
\colorlet{mycolor1}{green!50}
\colorlet{mycolor2}{orange!50}
\colorlet{mycolor3}{red!50}
\colorlet{mycolor4}{cyan!50}
\colorlet{mybackground1}{lightgray!50}

\definecolor{codegreen}{rgb}{0,0.6,0}
\definecolor{codegray}{rgb}{0.5,0.5,0.5}
\definecolor{codepurple}{rgb}{0.58,0,0.82}
\definecolor{backcolour}{rgb}{0.95,0.95,0.92}
\definecolor{lightGrey}{rgb}{0.9, 0.9, 0.9}
\definecolor{beaublue}{rgb}{0.74, 0.83, 0.9}
\definecolor{lightred}{RGB}{229, 220, 220}

\usepackage[capitalise]{cleveref}

\newcommand{\inlineimg}[1]{\raisebox{-0.25em}{\includegraphics[clip,height=1em]{#1}}}

\usepackage{float}
\newfloat{lstfloat}{htbp}{lop}
\floatname{lstfloat}{Listing}
\crefname{lstfloat}{Listing}{Listings}

\usepackage{listings}

\usepackage{colortbl}

\definecolor{codegreen}{rgb}{0,0.6,0}
\definecolor{codegray}{rgb}{0.5,0.5,0.5}
\definecolor{codepurple}{rgb}{0.58,0,0.82}
\definecolor{backcolour}{rgb}{0.95,0.95,0.92}

\lstdefinestyle{custom}{
    backgroundcolor=\color{backcolour},
    commentstyle=\color{codegreen},
    keywordstyle=\color{magenta},
    numberstyle=\tiny\color{codegray},
    stringstyle=\color{codepurple},
    basicstyle=\ttfamily\footnotesize,
    showstringspaces=false,
    breaklines,
    tabsize=2,
    numbers=left,
    columns=fullflexible,
    keepspaces=true,
    frame=lines,
    numbersep=2pt,
    escapechar=|,
    captionpos=b,
}

\lstdefinestyle{customc}{
    backgroundcolor=\color{backcolour},
    commentstyle=\color{codegreen},
    keywordstyle=\color{magenta},
    numberstyle=\tiny\color{codegray},
    stringstyle=\color{codepurple},
    basicstyle=\ttfamily\footnotesize,
    showstringspaces=false,
    breaklines,
    tabsize=2,
    numbers=left,
    columns=fullflexible,
    keepspaces=true,
    frame=lines,
    numbersep=2pt,
    escapechar=|,
    captionpos=b,
    language=c,
}

\lstdefinelanguage{wasm}{
    sensitive=true,
    otherkeywords={},
    morekeywords=[5]{i32, f32, i64, f64, add, const, sub, return, module, func, param, result, global, mut, global, export, import, memory, data, local, get, set, elem, table, call, call_indirect, type, segment, new, set_tag, free, tee, wrap_i64, pointer_sign, pointer_auth},
    keywordstyle={[5]\color{magenta}},
    numberstyle=\tiny\color{black},
    rulecolor=\color{black},
    morecomment=**[l][\itshape\color{codegreen}]{;;},
}

\lstdefinelanguage{llvm}{
    morecomment = [l]{;},
    morestring=[b]",
    sensitive = true,
    classoffset=0,
    morekeywords={
        define, declare, global, constant,
        internal, external, private,
        linkonce, linkonce_odr, weak, weak_odr, appending,
        common, extern_weak,
        thread_local, dllimport, dllexport,
        hidden, protected, default,
        except, deplibs,
        volatile, fastcc, coldcc, cc, ccc,
        x86_stdcallcc, x86_fastcallcc,
        ptx_kernel, ptx_device,
        signext, zeroext, inreg, sret, nounwind, noreturn,
        nocapture, byval, nest, readnone, readonly, noalias, uwtable,
        inlinehint, noinline, alwaysinline, optsize, ssp, sspreq,
        noredzone, noimplicitfloat, naked, alignstack,
        module, asm, align, tail, to,
        addrspace, section, alias, sideeffect, c, gc,
        target, datalayout, triple,
        blockaddress
    },
    classoffset=1, keywordstyle=\color{purple},
    morekeywords={
        fadd, sub, fsub, mul, fmul,
        sdiv, udiv, fdiv, srem, urem, frem,
        and, or, xor,
        icmp, fcmp,
        eq, ne, ugt, uge, ult, ule, sgt, sge, slt, sle,
        oeq, ogt, oge, olt, ole, one, ord, ueq, ugt, uge,
        ult, ule, une, uno,
        nuw, nsw, exact, inbounds,
        phi, call, select, shl, lshr, ashr, va_arg,
        trunc, zext, sext,
        fptrunc, fpext, fptoui, fptosi, uitofp, sitofp,
        ptrtoint, inttoptr, bitcast,
        ret, br, indirectbr, switch, invoke, unwind, unreachable,
        alloca, load, store, getelementptr,
        extractelement, insertelement, shufflevector,
        extractvalue, insertvalue,
        void, ptr, i8, i16, i32, i64, i128, double, float
    },
    alsoletter={\%},
    keywordsprefix={\%},
}

\lstdefinestyle{customwasm}{
    backgroundcolor=\color{backcolour},
    commentstyle=\color{codegreen},
    keywordstyle=\color{magenta},
    numberstyle=\tiny\color{codegray},
    stringstyle=\color{codepurple},
    basicstyle=\ttfamily\footnotesize,
    breaklines,
    tabsize=2,
    numbers=left,
    columns=fullflexible,
    keepspaces=true,
    frame=lines,
    numbersep=2pt,
    escapechar=|,
    captionpos=b,
    language=wasm,
    keywords={}
}

\usepackage{subcaption}
\usepackage{mathtools}
\usepackage{algorithmic}
\usepackage{multirow}

\usepackage{acronym}

\acrodef{ASAN}[ASan]{Address Sanitizer}
\acrodef{ASLR}[ASLR]{Address Space Layout Randomization}
\acrodef{CFI}[CFI]{Control flow integrity}
\acrodef{CHERI}[CHERI]{Capability Hardware Enhanced RISC Instructions}
\acrodef{CLIF}[CLIF]{Cranelift IR}
\acrodef{CPI}[CPI]{code-pointer integrity}
\acrodef{CPS}[CPS]{code-pointer separation}
\acrodef{GEP}[GEP]{\texttt{getelementptr}}
\acrodef{HWASAN}[HWASan]{Hardware-Assisted Address Sanitizer}
\acrodef{IR}[IR]{intermediate representation}
\acrodef{ISA}[ISA]{instruction set architecture}
\acrodef{MMU}[MMU]{Memory Management Unit}
\acrodef{MTE}[MTE]{Memory Tagging Extension}
\acrodef{PAC}[PAC]{Pointer Authentication}
\acrodef{ROP}[ROP]{return-oriented programming}
\acrodef{rss}[rss]{resident set size}
\acrodef{TBI}[TBI]{Top Byte Ignore}
\acrodef{WASI}[WASI]{WebAssembly System Interface}
\acrodef{WASM}[WASM]{WebAssembly}

\usepackage{xurl}

\newcommand{\projectname}{\textsc{Cage}\xspace}

\usepackage[inline]{enumitem}
\setlist{leftmargin=16pt,noitemsep,topsep=0pt,parsep=0pt,partopsep=0pt}

\usepackage{setspace}
\usepackage[margin=5pt,font={stretch=0.9}]{caption}

\begin{document}

\title{\projectname: Hard\-ware-Accelerated Safe WebAssembly}

\author{Martin Fink}
\email{martin.fink@cit.tum.de}
\orcid{0000-0002-3280-8974}
\affiliation{%
    \institution{Technical University of Munich}
    \city{Munich}
    \country{Germany}
}

\author{Dimitrios Stavrakakis}
\email{dimitrios.stavrakakis@tum.de}
\orcid{0000-0002-3667-3763}
\affiliation{%
    \institution{Technical University of Munich}
    \city{Munich}
    \country{Germany}
}

\author{Dennis Sprokholt}
\email{d.g.sprokholt@tudelft.nl}
\orcid{0000-0002-2132-7315}
\affiliation{%
    \institution{Delft University of Technology}
    \city{Delft}
    \country{The Netherlands}
}

\author{Soham Chak\-ra\-borty}
\email{S.S.Chakraborty@tudelft.nl}
\orcid{0000-0002-4454-2050}
\affiliation{%
    \institution{Delft University of Technology}
    \city{Delft}
    \country{The Netherlands}
}

\author{Jan-Erik Ekberg}
\email{jan.erik.ekberg@huawei.com}
\orcid{0009-0007-5432-6128}
\affiliation{%
    \institution{Huawei Technologies}
    \city{Helsinki}
    \country{Finland}
}

\author{Pramod Bhatotia}
\email{pramod.bhatotia@tum.de}
\orcid{0000-0002-3220-5735}
\affiliation{%
    \institution{Technical University of Munich}
    \city{Munich}
    \country{Germany}
}

\renewcommand{\shortauthors}{M. Fink, D. Stavrakakis, D. Sprokholt, S. Chakraborty, J.-E. Ekberg, P. Bhatotia}

\begin{abstract}
\ac{WASM} is an immensely versatile and increasingly popular compilation target. It executes applications written in several languages (e.g., C/C++) with near-native performance in various domains (e.g., mobile, edge, cloud).
Despite \ac{WASM}'s sandboxing feature, which isolates applications from other instances and the host platform, \ac{WASM} does not inherently provide any memory safety guarantees for applications written in low-level, unsafe languages.

To this end, we propose \projectname{}, a hardware-accelerated toolchain for \ac{WASM} that supports \emph{unmodified} applications compiled to \ac{WASM}
and utilizes diverse Arm hardware features aiming to enrich the memory safety properties of \ac{WASM}.
Precisely, \projectname{} leverages Arm's \ac{MTE} to \emph{(i)}~provide spatial and temporal memory safety for heap and stack allocations and \emph{(ii)}~improve the performance of \ac{WASM}'s sandboxing mechanism.
\projectname{} further employs Arm's \ac{PAC} to prevent leaked pointers from being reused by other \ac{WASM} instances, thus enhancing \ac{WASM}'s security properties.

We implement our system based on 64-bit \ac{WASM}.
We provide a \ac{WASM} compiler and runtime with support for Arm's \ac{MTE} and \ac{PAC}.
On top of that, \projectname{}'s LLVM-based compiler toolchain transforms unmodified applications to provide spatial and temporal memory safety for stack and heap allocations and prevent function pointer reuse.
Our evaluation on real hardware shows that \projectname{} incurs minimal runtime~($<5.8\,\%$) and memory~($<5.3\,\%$) overheads and can improve the performance of \ac{WASM}'s sandboxing mechanism, achieving a speedup of over $5.1\,\%$, while offering efficient memory safety guarantees.
\end{abstract}




\maketitle

\section{Introduction}
\label{sec:intro}

WebAssembly (WASM)~\cite{haas2017bringing} has been gaining prominence as a versatile compilation target~\cite{musch2019new}.
\ac{WASM} allows for deploying and executing native applications written in a variety of languages, such as C/C++ and Rust, in a wide spectrum of environments (e.g., Web, Edge Cloud, IoT devices)~\cite{wasm_use_cases}, while achieving near-native performance.
In principle, \ac{WASM} enables the compilation of high-level languages (e.g., C/C++) to its bytecode format, which can then be seamlessly compiled to native machine code based on the targeted underlying architecture, which justifies its increasing adoption.

A core design principle of \ac{WASM} is the \emph{sandboxing} of untrusted code~\cite{wasm_sandbox},
providing a safety property of Software Fault Isolation (SFI)~\cite{wahbe1993efficient}.
Each \ac{WASM} application is confined in its isolated address space and has no access rights beyond that.
Thus, \ac{WASM} protects the host and other guest \ac{WASM} instances from potentially malicious or buggy code.

However, applications compiled to \ac{WASM} are still vulnerable to memory safety issues, such as buffer overflows or dangling pointers, within an application's memory space, despite \ac{WASM}'s sandboxing~\cite{lehmann2020everything}.
Such issues allow attackers to manipulate the memory space of a \ac{WASM} instance, corrupting or leaking sensitive data or manipulating the control flow.
This limitation becomes particularly evident when compiling applications written in memory-unsafe languages, e.g., C/C++.
Importantly, numerous CVEs remain exploitable when compiled to \ac{WASM} ($\S$\ref{sec:motivation}).

Unfortunately, existing memory safety solutions, designed for applications compiled to native machine code, cannot be directly applied for \ac{WASM} due to several factors:
\begin{itemize}[noitemsep, leftmargin=*]
    \item[--] \textbf{Memory safe languages}, such as Rust or OCaml, tackle memory safety at the language level by encoding certain properties in their type system or providing safe abstractions, e.g., in their standard libraries.
    However, it is not feasible to port all applications or libraries written in C/C++ to languages with such guarantees.
    \item[--] \textbf{Trip-wire-based} memory safety approaches~\cite{serebryany2012addresssanitizer,serebryany2023gwp,bozdougan2022safepm,nethercote2007valgrind,lightweight_bounds_checking,shadow_stack} do not apply to \ac{WASM} without altering the \ac{WASM} memory layout properties, as these approaches rely on shadow memory or guard pages.
    \item[--] \textbf{Pointer- or object-based} solutions~\cite{nagarakatte2009softbound, delta, sgx-bounds, akritidis2009baggy, nethercote2007valgrind, michael2023mswasm,lowfat,safecode,stack_bounds} can be adapted to \ac{WASM}, but they either modify the pointer layout and size, incur significant runtime overheads, or rely on custom hardware~\cite{watson_cheri_2020}, thus, hindering production deployment in common \ac{WASM} environments.
\end{itemize}

\noindent Hence, we aim to answer the question:
\emph{How can we design a practical system to provide memory safety properties for unmodified applications compiled to \ac{WASM} without changing its linear memory model and introducing minimal overheads to allow for deployment in production environments?}

Our key insight is to leverage modern commodity hardware extensions by designing abstractions for \ac{WASM} that can be directly used by the compilers.
Precisely, Arm recently introduced new \acs{ISA} extensions, namely \ac{PAC}~\cite{Qualcomm2017PointerAuth} and \ac{MTE}~\cite{ARM2019MTE}, that can serve as a base to design practical, high-performance solutions against memory safety issues.

To this end, we propose \projectname, a hardware-accelerated \ac{WASM} toolchain that leverages Arm's \ac{MTE} and \ac{PAC} hardware extensions to provide both spatial and temporal memory safety issues for \emph{unmodified} C/C++ programs. \projectname{} further hardens applications against control flow highjacking in the form of function pointer reuse between \ac{WASM} instances.
Additionally, \projectname improves the performance of the sandboxing mechanism for 64-bit \ac{WASM} by offloading the bounds checks to \ac{MTE} hardware.
We design \projectname so it can fall back to software-based implementations on devices lacking the relevant hardware.

To the best of our knowledge, \projectname{} is the \emph{first} practical, comprehensive, and efficient solution for unmodified C/C++ programs that can be deployed on commodity hardware and provides strong memory safety guarantees for \ac{WASM} instances running memory-unsafe code on Arm platforms.

Altogether, \projectname{} makes the following contributions:
\begin{itemize}[noitemsep, leftmargin=*]
    \item[--]{\textbf{WebAssembly extension:} A minimal and generic extension to the WebAssembly specification to provide memory safety guarantees and is deployable on every platform, regardless of the availability of specialized hardware.}
    \item[--]{\textbf{Compiler toolchain:} A compiler toolchain that transparently hardens unmodified programs to enforce spatial and temporal memory safety for stack and heap allocations and prevent unintended function pointer reuse.}
    \item[--]{\textbf{\ac{WASM} runtime:} A hardware-accelerated \ac{WASM} compiler and runtime, leveraging Arm's \ac{MTE} and \ac{PAC} with minimal overhead that can be deployed in production.}
    \item[--]{\textbf{Evaluation on real hardware:} Evaluation and security analysis of \projectname{}'s implementation on commercially available Arm hardware. Our evaluation is structured around performance and memory overheads and is accompanied by an extensive analysis of \ac{MTE} and \ac{PAC} performance as implemented on real hardware.}
\end{itemize}

\noindent
We implement our \projectname{} prototype on top of 64-bit \ac{WASM}, consisting of a compiler toolchain based on LLVM and a \ac{WASM} compiler and runtime based on wasmtime incorporating support for Arm's \ac{MTE} and \ac{PAC}.
We conduct an extensive analysis of \projectname{}'s security guarantees and evaluate \projectname{}'s performance using the PolyBench/C~\cite{polybenchc} benchmark suite.
We further analyze Arm's \ac{MTE} and \ac{PAC} performance as implemented on production hardware through a set of microbenchmarks.
Our evaluation on real Arm hardware shows that \projectname{} provides its memory safety properties while incurring minimal runtime ($<5.6\,\%$) and memory ($<5.3\,\%$) overheads and is capable of significantly improving the performance of \ac{WASM}'s sandboxing feature, achieving a speedup of over $5.1\,\%$.

\section{Background}
\label{sec:background}

\subsection{WebAssembly}
\label{subsec:wasm}

\begin{figure}[t]
    \centering
    \includegraphics[scale=1]{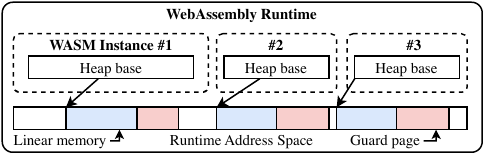}
    \caption{\ac{WASM}'s linear memory model.}
    \Description{\ac{WASM}'s linear memory model.}
    \label{fig:wasm-linear-memory}
\end{figure}

\Acl{WASM}~\cite{haas2017bringing} is a versatile, high-performance compilation target, initially designed as an alternative to JavaScript.
Its goal is to execute applications written in any language with near-native performance regardless of the hardware and software stack of the hosting platform.
Several high-level languages (e.g., C/C++, Rust) can be compiled to \Ac{WASM}'s bytecode format.
It is then lowered to the appropriate native machine code, depending on the underlying system architecture.
This feature expands the usability of \Ac{WASM} to various other domains~\cite{wasm_use_cases}, such as Function-as-a-Service (FaaS) workloads or even as an alternative to Linux containers~\cite{docker_wasm}.

\Ac{WASM} employs a linear memory model (see \cref{fig:wasm-linear-memory}).
Thus, applications manage their memory without requiring unnatural idioms, and \ac{WASM} runtimes can efficiently map the \ac{WASM} instance memory directly to host memory.
Importantly, \ac{WASM}'s design does not allow unstructured control flow.
\Ac{WASM} uses indices into type- and bounds-checked tables instead of raw function pointers to make indirect function calls, while, for jumps, \ac{WASM} provides a set of well-defined control flow constructs.
Additionally, \ac{WASM} does not expose registers but operates on a verified, well-typed stack to ensure compatibility with diverse compilation targets that offer different sets of registers.

\ac{WASM} also provides \emph{sandboxing} for programs.
The \ac{WASM} runtime must ensure that each instance can only access memory within the bounds of its accessible linear memory.
This is typically achieved using either
\emph{(i)}~\emph{explicit bounds checks} that are inserted before every memory access, validating that it lies within the \ac{WASM} instance's memory range
or 
\emph{(ii)}~\emph{guard pages}, where 4\,GiB of virtual memory is mapped and pages beyond the \ac{WASM} instance's memory is marked as inaccessible, and any access in them results in a segmentation fault.
This only works for 32\,bit pointers, which don't allow accessing memory beyond 4\,GiB.
Typically, switching to 64-bit \ac{WASM} entails switching to the more expensive approach \emph{(i)} with explicit bounds checks.

\begin{figure}[t]
    \centering
    \includegraphics[scale=1]{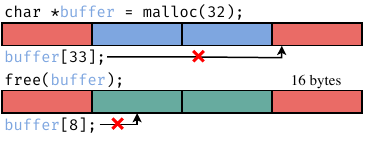}
    \caption{Example of a heap allocation protected by \ac{MTE}. After allocation, the pointer and allocation are tagged with~\inlineimg{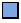}, while the surrounding memory is tagged with~\inlineimg{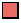}. When accessing memory, the hardware catches out-of-bounds errors ($\inlineimg{./figures/build/color-blue} \neq \inlineimg{./figures/build/color-red}$). When freeing memory, the memory region is retagged with \inlineimg{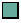}. This prevents use-after-free errors ($\inlineimg{./figures/build/color-blue} \neq \inlineimg{./figures/build/color-green}$).}
    \label{fig:mte}
\end{figure}

\subsection{Memory Safety}
\label{subsec:memory-safety-wasm}

Applications written in low-level, memory-unsafe languages (e.g., C/C++) are prone to memory safety bugs that enable a whole class of attacks on a vulnerable or buggy program~\cite{szekeres2013sok}.
Several studies have shown that in large software projects, memory safety bugs make up between 70\,\% and 75\,\% of their security vulnerabilities~\cite{chromium_memory_safety,microsoft_memory_safety,android_memory_safety}.
These bugs are classified as \emph{spatial} or \emph{temporal}.
Spatial memory safety errors occur when a memory access is performed beyond the allocated boundaries of a memory object (e.g., buffer overflows), while temporal memory safety bugs refer to accesses to memory regions before they are allocated or after their release (e.g., dangling pointers).
These vulnerabilities can be exploited by attackers to overwrite data or redirect control flow, for instance, by manipulating the return address saved on the stack to create \ac{ROP} chains, chaining existing snippets of code together to create attacks.

To this end, several memory safety solutions have been proposed, which can be divided into three major categories: \emph{(i)~trip-wire-based approaches}~\cite{serebryany2012addresssanitizer, serebryany2023gwp, bozdougan2022safepm, nethercote2007valgrind, lightweight_bounds_checking, shadow_stack} that employ guard zones around memory objects and allocate designated memory regions, namely shadow memory, that determine whether a memory location is accessible or not, \emph{(ii)~object-based approaches}~\cite{akritidis2009baggy, lowfat, safecode, stack_bounds} that track memory safety metadata on a per-object level and ensure memory safety for pointers with respect to an object, and \emph{(iii)~pointer-based approaches}~\cite{nagarakatte2009softbound, necula2002ccured, jim2002cyclone, delta, Intel2013MPX, mpx-explained, woodruff2014cheri, sgx-bounds} that keep track of object bounds by either storing them in the pointer itself (e.g., via fat pointers) or in external data structures.
Additionally, specialized solutions~\cite{abadi2009control, kuznetzov2018code, liljestrand2019pac} have also been proposed to mitigate control-flow attacks that occur via memory safety bugs and either analyze applications to enforce valid control flow paths or enforce memory safety for code pointers using hardware- or software-based techniques.

\noindent
Memory tagging approaches~\cite{serebryany2018memory, ARM2019MTE} combine aspects from object- and pointer-based approaches. They typically associate metadata stored in the unused bits of a pointer with memory objects by assigning tags to allocated memory regions and performing checks at runtime.

\begin{figure}[t]
    \centering
    \includegraphics[scale=1]{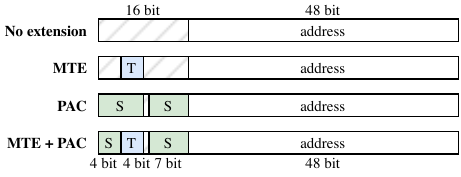}
    \caption{Pointer layout on aarch64 in Linux with and without \ac{MTE} and \ac{PAC} enabled.}
    \Description{Pointer layout on aarch64 in Linux with and without \ac{MTE} and \ac{PAC} enabled.}
    \label{fig:aarch64-pointer}
\end{figure}

\subsection{Memory Safety Hardware Extensions}
\label{subsec:memory-safety-hardware-extensions}

As an alternative to flexible yet slow software-based memory safety solutions, CPU designers develop hardware extensions~\cite{ARM2019MTE, Qualcomm2017PointerAuth, Intel2013MPX, mpx-explained} to serve as a foundation for efficient memory safety.
They provide security primitives to fortify applications with memory safety properties while having minimal memory and performance overheads, making them suitable for production deployment.

\cref{fig:aarch64-pointer} presents an example layout of a pointer on \texttt{aarch64}, the 64-bit variant of the ArmV8 \acs{ISA}~\cite{ARMA2024Arch64}.
Only 48 out of the available 64\ bits are used to address memory. The remaining bits are set to 0 or 1 to differentiate between kernel and userspace.
Hardware extensions such as \ac{TBI}, \ac{MTE} ($\S$\ref{subsubsec:mte}), or \ac{PAC} ($\S$\ref{subsubsec:pac}) leverage those unused bits for storing metadata.

\myparagraph{Memory Tagging Extension (MTE)}
\label{subsubsec:mte}
ARMs \Ac{MTE}
provides a building block for spatial and temporal memory safety~\cite{ARM2019MTE}.
\Ac{MTE} implements a lock-and-key mechanism where memory regions can be tagged with one of 16 distinct tags, and memory accesses are only allowed using pointers with the corresponding tags.
The locking mechanism stores a 4-bit tag in bits 56--59 of an address. 
Accordingly, a tag is assigned to memory with a granularity of 16\,bytes.

\cref{fig:mte} presents how \Ac{MTE} can provide \emph{spatial} as well as \emph{temporal} memory safety.
Precisely, \Ac{MTE} can ensure spatial memory safety by assigning different tags to adjacent memory regions, while it also can offer temporal memory safety by retagging freed memory regions.

\ac{MTE} currently can be set in one the following modes: \emph{(i)}~\emph{disabled}, where no tag checks are performed, \emph{(ii)}~\emph{synchronous}, where a tag mismatch immediately triggers a fault disallowing the read/write at the affected memory location, \emph{(iii)} \emph{asynchronous}, where a tag mismatch does not cause a fault directly but sets a CPU flag that is checked at the next context switch, thus allowing for a potential read/write at the affected memory location by the triggering command, and \emph{(iv) asymmetric}, where reads are checked asynchronously and writes are checked synchronously.

\newcolumntype{C}{>{\centering\arraybackslash}p{2em}}
\newcolumntype{P}[1]{>{\centering\arraybackslash}p{#1}}
\begin{table}[t]
    \centering
    \small
    \caption{MTE and PAC instruction throughput (instructions per cycle, higher is better) and latencies (cycles, lower is better). We only show PAC instructions using the Data A-key (\texttt{da}).}
    \label{tab:instruction-latencies}
    \begin{tabular}{P{10mm} || P{6mm} | P{6mm} | P{6mm} | P{6mm} | P{6mm} | P{6mm} }
        \multirow{2}{*}{\textbf{Inst}} & \multicolumn{2}{c|}{\textbf{Cortex-X3}} & \multicolumn{2}{c|}{\textbf{Cortex-A715}} & \multicolumn{2}{c}{\textbf{Cortex-A510}} \\
        & Tp   & Lat  & Tp   & Lat  & Tp   & Lat  \\ \hline
        & \multicolumn{6}{c}{\textbf{MTE}} \\ \hline
        \texttt{irg}    & 1.34 & 1.99 & 1.00 & 2.00 & 0.50 & 3.00 \\
        \texttt{addg}   & 2.01 & 1.99 & 3.81 & 1.00 & 2.22 & 2.00 \\
        \texttt{subg}   & 2.01 & 1.99 & 3.81 & 1.00 & 2.22 & 2.00 \\
        \texttt{subp}   & 3.49 & 0.99 & 3.81 & 1.00 & 2.50 & 2.00 \\
        \texttt{subps}  & 2.88 & 0.99 & 3.80 & 1.00 & 2.50 & 2.00 \\
        \texttt{stg}    & 1.00 & --   & 1.81 & --   & 1.00 & --   \\
        \texttt{st2g}   & 1.00 & --   & 1.84 & --   & 0.46 & --   \\
        \texttt{stzg}   & 1.00 & --   & 1.84 & --   & 0.98 & --   \\
        \texttt{st2zg}  & 0.34 & --   & 1.79 & --   & 0.45 & --   \\
        \texttt{stgp}   & 1.00 & --   & 1.69 & --   & 0.98 & --   \\
        \texttt{ldg}    & 2.92 & --   & 1.91 & --   & 0.93 & --   \\ \hline
        & \multicolumn{6}{c}{\textbf{PAC}} \\ \hline
        \texttt{pacdza} & 1.01 & 4.97 & 1.51 & 5.00 & 0.20 & 4.99 \\
        \texttt{pacda}  & 1.01 & 4.97 & 1.42 & 5.00 & 0.20 & 5.00 \\
        \texttt{autdza} & 1.01 & 4.97 & 1.51 & 5.00 & 0.20 & 7.99 \\
        \texttt{autda}  & 1.01 & 4.97 & 1.43 & 5.00 & 0.20 & 7.99 \\
        \texttt{xpacd}  & 1.01 & 1.99 & 1.56 & 2.00 & 0.20 & 4.99 \\
    \end{tabular}
\end{table}

\myparagraph{Pointer Authentication (PAC)}
\label{subsubsec:pac}
\Ac{PAC}~\cite{Qualcomm2017PointerAuth} introduces hardware primitives to prevent attackers from forging pointers.
\ac{PAC} places a 7 to 16\,bit signature in the upper bits of pointers, with the exact layout being dependent on the operating system, the underlying hardware, and other factors (e.g. if \ac{MTE} is enabled).
Signatures are created using the pointer value, a secret key placed in an inaccessible register, and a user-defined value (modifier).
Signed pointers cannot be used directly to access memory; they must be authenticated.
Authenticating a pointer consists of validating the signature and stripping out the signature if the validation is successful, thus producing a valid pointer. In case of a failed authentication, \ac{PAC} can either produce a pointer that will trap on memory access or trap immediately.
This behavior depends on whether FEAT\_FPAC is implemented~\cite{ARMA2024Arch64}.

\ac{MTE} and \ac{PAC} can be combined at the cost of bits available for the \ac{PAC} signature.
The exact layout of the \ac{PAC} signature varies depending on the system.
On Linux, bits 56--59 are used for \ac{MTE} while bits 63--60 and 54--49 are used for \ac{PAC}, as shown in \cref{fig:aarch64-pointer}.
The remaining bit 55 differentiates between kernel- and user-space addresses.

\begin{figure}[t]
    \centering
    \includegraphics[width=0.9\columnwidth]{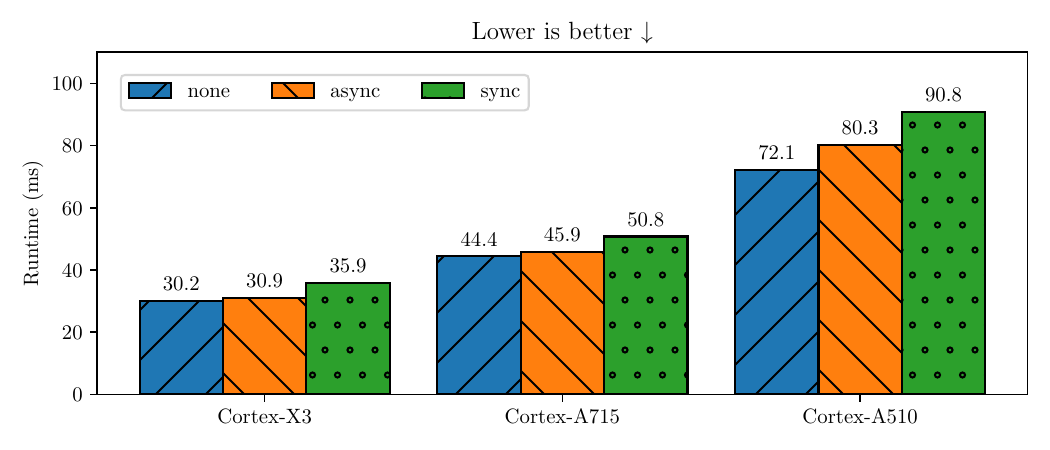}
    \caption{Performance overhead of \ac{MTE} sync and async mode for writing 128\,MiB of memory. See $\S$\ref{subsec:experimental-setup} for details on the experimental setup.}
    \label{fig:sync-async-performance}
\end{figure}

\myparagraph{Architectural performance analysis}
We evaluate the performance characteristics of \ac{MTE} as implemented on the Tensor G3 chip, such as throughput and latency of the individual instructions in \cref{tab:instruction-latencies}.
We run microbenchmarks executing $10^{10}$ instructions in an unrolled loop.
To measure throughput, we execute the instructions without any data dependencies; to measure latency, we force a data dependency between subsequent instructions.
For instructions storing and loading memory tags, we only measure throughput.

We further measure the raw overhead of enabling \ac{MTE}. We perform a 128\,MiB \texttt{memset} with \ac{MTE} disabled, synchronous, and asynchronous mode.
We perform each run with a clean cache.
In \cref{fig:sync-async-performance}, we observe that with synchronous \ac{MTE}, \texttt{memset} is 19.1\%, 14.4\%, and 29.9\% slower on the respective cores compared to the baseline with \ac{MTE} disabled.
Asynchronous \ac{MTE} gets closer to the baseline with an overhead of 2.6\%, 3.3\%, and 11.3\%, respectively.

\section{Motivation}
\label{sec:motivation}

WebAssembly is inherently protected by design against a wide range of attack vectors, such as jumping to arbitrary addresses or injecting shellcode.
Despite its protection mechanisms, it is shown that WebAssembly is still susceptible to attacks originating from memory safety issues, such as buffer overflows or dangling pointer accesses~\cite{lehmann2020everything}.

Importantly, \ac{WASM} compilers place data in the linear memory of the \ac{WASM} instance, where both read and write permissions are granted to the executed code.
The lack of read-only memory regions and the ability to map arbitrary pages in \ac{WASM} prevents measures, such as \ac{ASLR} or guard pages, from being applied.
Thus, in case of a successful exploit of a memory safety bug, an attacker can overwrite application data.
\Cref{tab:cves} presents a set of exploits that were discovered outside the scope of \ac{WASM} but serve as examples that showcase potential memory safety errors when the vulnerable applications run in a \ac{WASM} instance.

\begin{table}[t]
    \small
    \centering
    \caption{An exemplary list of memory safety errors, their underlying cause, and the level of their mitigation.}
    \label{tab:cves}
    \begin{tabular}{c|c|c}
        \textbf{CVE} & \textbf{Cause} & \textbf{Mitigated in WASM} \\\hline
        CVE-2023-4863 & Out-of-bounds & No \\
        CVE-2014-0160 & Out-of-bounds & No \\
        CVE-2021-3999 & Out-of-bounds & Partially\footnotemark\label{footnote} \\
        CVE-2018-14550 & Out-of-bounds & No \\
        CVE-2021-22940 & Use-after-free & No \\
        CVE-2021-33574 & Use-after-free & No \\
        CVE-2020-1752 & Use-after-free & No \\
        CVE-2019-11932 & Double-free & Partially\footnotemark[\value{footnote}]{} \\
    \end{tabular}
\end{table}
\footnotetext[\value{footnote}]{Although the ROP chain is mitigated, memory corruption is still possible.}

\begin{figure*}[t]
    \centering
    \includegraphics{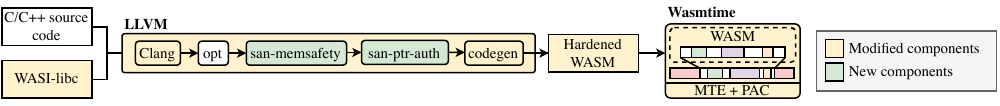}

    \caption{Overview of \projectname{}'s components and instrumentation pipeline: Unmodified C/C++ code is compiled along with our modified WASI-libc. After optimizations, two sanitizer passes for memory safety and pointer authentication run, inserting new \ac{WASM} instructions. Wasmtime processes this hardened \ac{WASM} and emits \ac{MTE} and \ac{PAC}.}
    \label{fig:system-overview}
\end{figure*}

Additionally, in \ac{WASM}, a degree of control flow manipulation is possible.
Precisely, function pointers can be overwritten with pointers to other functions that share the same signature.
However, it can only occur with functions that are present in the function table, i.e., functions that are targets of virtual function calls, such as virtual functions in C++.
This results in a similar property as \ac{CPS}~\cite{kuznetzov2018code}.
\cref{lst:vulnerable-overflow-pac} showcases such a vulnerability, as an attacker can overwrite a function pointer and redirect an indirect call to another function.
When WASM is used in a WebOS-like scenario~\cite{kwast_os, redshirt, wasmachine}, i.e., running multiple instances in a single process, leaking function pointers in one program is even more vulnerable, especially if instances share a common library.

\myparagraph{WebAssembly sandboxing}
\Ac{WASM} engines use various techniques to protect their sandboxes against malicious code.
While virtual memory and guard pages are preferred for performance reasons, some settings (e.g., 64-bit \ac{WASM}) necessitate software-based bounds checks, which come at a higher performance cost.
Based on our evaluation, switching from 32-bit to 64-bit \ac{WASM} results in a roughly 6--8\,\% overhead on out-of-order CPUs, which can speculate through bounds checks, and 52\,\% overhead on in-order CPUs ($\S$~\ref{subsec:performance-overheads}).
The measurements on out-of-order CPUs align with previous works~\cite{szewczyk2022leaps}.
The fallback to software-based bounds checks is especially painful when running on low-power in-order cores using 64-bit \ac{WASM} or in environments without an OS, such as embedded devices.
Additionally, software bounds checks or the guard pages technique may suffer from implementation bugs and must be protected against spectre-style attacks~\cite{kocher2020spectre}.
An example is CVE-2023-26489~\cite{CVE-2023-26489}, where an erroneous code lowering rule allowed malicious \ac{WASM} instances to access memory outside the sandbox.

\Ac{WASM} is, despite limitations such as high overhead when running 64-bit \ac{WASM} or the lack of a practical and low-overhead solution to prevent memory safety issues in programs written in C/C++, steadily growing in adoption~\cite{wasm_use_cases}. \\
Recently, manufacturers have started shipping hardware supporting both PAC and MTE, with the Google Pixel 8 being the first commercially available device to feature both.
These extensions offer strong security guarantees with very low overhead, as shown in \cref{tab:instruction-latencies} and \cref{fig:sync-async-performance}.

\begin{lstfloat}[t]
\footnotesize
    \centering
    \begin{lstlisting}[frame=h,style=customc,label={lst:vulnerable-overflow-pac-inner}]
struct VTable { void (*f)(); void (*g)(); };
void vulnerable(char *input) {
  struct VTable vtable = {.f = foo, .g = bar};
  char buf[16];
  strcpy(buf, input);
  vtable.f();
}\end{lstlisting}
    \caption{Vulnerable overflow allowing attackers to redirect control flow to call \texttt{foo} instead of \texttt{bar}.}
    \label{lst:vulnerable-overflow-pac}
\end{lstfloat}

\myparagraph{\projectname{}}
To tackle the aforementioned memory safety issues, we design \projectname{}, an extension to \ac{WASM} that efficiently prevents memory safety and function-pointer reuse exploits by leveraging Arm's \ac{MTE} and \ac{PAC}.
On top of that, \projectname{} implements hardware-based sandboxing using \ac{MTE} to circumvent the overhead of software-based checks and prevent vulnerabilities, e.g., CVE-2023-26489.

\section{Design}
\label{sec:design}

\projectname{} consists of a \ac{WASM} extension to provide memory safety guarantees within a \ac{WASM} instance and improve the performance of the sandboxing mechanism.
Its core design principles are to be \emph{(i)}~minimally invasive and \emph{(ii)}~applicable on diverse platforms using various approaches, including hardware extensions, such as \ac{MTE} and \ac{PAC}, software-based techniques, or hybrid solutions similar to \acs{HWASAN}~\cite{serebryany2018memory}.

Figure~\ref{fig:system-overview} presents an overview of \projectname{}.
Precisely, unmodified C/C++ source code is compiled using the LLVM toolchain~\cite{lattner2004llvm}, where \projectname{} includes a sanitizer that instruments stack allocations and function pointers, along with a modified libc based on \acs{WASI}~\cite{wasi} that protects heap allocations.
LLVM's backend generates the hardened \ac{WASM} binaries that can be executed in wasmtime~\cite{wasmtime}, which implements our extension using \ac{MTE} and \ac{PAC}.

\subsection{System Model}
\label{subsec:system-model}

\begin{figure}[t]
    \centering
    \begin{subfigure}[t]{0.2\textwidth}
        \centering
        \vskip 0pt
        \includegraphics{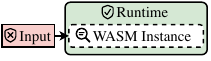}
        \caption{Internal memory safety.}
        \label{fig:internal-mem-safety}
    \end{subfigure}
    \hfill
    \begin{subfigure}[t]{0.2\textwidth}
        \centering
        \vskip 0pt
        \includegraphics{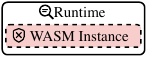}
        \caption{External memory safety.}
        \label{fig:external-mem-safety}
    \end{subfigure}
    \caption{Internal and external memory safety in \ac{WASM}.}
    \label{fig:threat-model}
\end{figure}

\myparagraph{Threat model}
\label{par:threat-model}
\Cref{fig:threat-model} highlights two aspects of memory safety, present in \ac{WASM}, namely \emph{internal} and \emph{external} memory safety.
We depict trusted components (\inlineimg{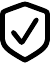}) in green and untrusted components (\inlineimg{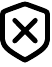}) in red.
We further annotate the hardened component by \projectname{} in each of these models (\inlineimg{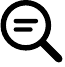}).

The \textit{internal memory safety} model (\cref{fig:internal-mem-safety}) mirrors the threat environment of a standard non-\ac{WASM} program.
In this case, the application in the sandbox and its runtime, including the compiler, are considered trusted and assumed to be bug-free.
Untrusted input (e.g., network data) originates outside the sandbox and can be controlled by an attacker. This implies that common memory safety bugs, such as buffer overflows or use-after-free, can be exploited to tamper with the \ac{WASM} memory.
WebAssembly's design inherently mitigates some threats common in non-\ac{WASM} environments ($\S$~\ref{subsec:wasm}), so we do not consider \ac{ROP}-style attacks or those relying on unstructured control flow.

The \textit{external memory safety} model (\cref{fig:external-mem-safety}) refers to the security of the sandbox.
Threats originate from running untrusted programs, which may be adversarial or contain bugs. 
Typical attacks include sandbox escapes, where an attacker attempts to break out of the sandbox's restrictions and access host resources, or side-channel attacks, where attackers exploit timing differences or resource usage patterns to infer sensitive information.
In this setting, we assume that the operating system and the underlying architecture do not have exploitable bugs.
Importantly, we do not make assumptions about potential spectre-like~\cite{kocher2020spectre} attacks. We do not protect against protected against side-channel attacks.

\myparagraph{Programming model}
\projectname{} currently supports unmodified applications written in C/C++ that target 64-bit \ac{WASM}.
Note that our toolchain is not bound to any language and can handle other languages compiled to LLVM in the future.
\projectname{} provides spatial and temporal heap safety to applications that use its adapted WASI-libc that comes with a hardened allocator.
For applications using their own allocator, we expose \projectname{}'s memory safety primitives to C, enabling programmers to implement the same security guarantees.
Lastly, \projectname{} reserves the unused upper 16\,bits of 64-bit pointers to place memory safety metadata.

\myparagraph{Deployment model}
\projectname{}'s prototype is highly optimized to use Arm's hardware extensions to minimize the runtime overheads.
Therefore, its primary deployment target is Arm CPUs with Arm \ac{PAC} and \ac{MTE} extensions.
However, \projectname{} can also be deployed on any platform, regardless of the underlying hardware, where the respective memory safety protection mechanisms have an equivalent software fallback, resulting in higher performance overheads.

\begin{figure}[t]
    \small
    \begin{align*}
        \text{(new instructions) } e \Coloneqq&\ \textbf{segment.new}\ o \\
        \mid&\ \textbf{segment.set\_tag}\ o \\
        \mid&\ \textbf{segment.free}\ o \\
        \mid&\ \textbf{i64.pointer\_sign} \\
        \mid&\ \textbf{i64.pointer\_authenticate}
    \end{align*}
    \caption{\projectname{}'s new instructions: Segment instructions take a constant unsigned offset $o$, which allows compilers to fold in constant offsets when manipulating segments.}
    \label{fig:new-inst}
\end{figure}

\subsection{WebAssembly Extension}
\label{subsec:wasm-extension}

\projectname{} essentially is an extension to WebAssembly that introduces primitives that can be used by a modified standard library or sanitizers to provide memory safety guarantees for selected memory allocations.
It builds on \texttt{wasm64}~\cite{memory64}, the 64-bit variant of WebAssembly.
\texttt{Wasm64} extends pointers within a \ac{WASM} instance to 64\,bits. Out of those, only 48 are used to index memory.
This allows \projectname{} to utilize the remaining 16\,bits to store its required memory safety metadata.

\myparagraph{Memory safety}
\projectname{} introduces the notion of \textit{segments} and \textit{tagged pointers} in the context of \ac{WASM}.
It further features five new instructions, shown in \cref{fig:new-inst}, that allow the creation of segments and the derivation of tagged pointers from raw pointers.
The \projectname{} pointers carry provenance and can only access the segment they were created with.
Conversely, segments can only be accessed by the tagged pointers created with them. 

At an instance startup, the linear memory consists of a single segment that can be accessed via untagged indices, allowing unmodified code to run under our new semantics without modifications.
This design choice allows the gradual integration of safety primitives into specific parts of WebAssembly applications where enhanced security is required.
Thus, \projectname{} can achieve security properties similar to those of specialized hardware designed mainly for memory safety~\cite{woodruff2014cheri} by combining the concept of segments with Arm's hardware extensions (e.g., \ac{MTE}).

\begin{figure}[t]
    \centering
    \begin{subfigure}{\columnwidth}
        \centering
        \includegraphics{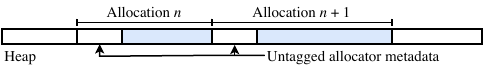}
        \caption{Heap layout with untagged allocator metadata slots, ensuring no tag collisions for adjacent allocations.}
        \label{fig:heap-layout}
    \end{subfigure}
    \\
    \begin{subfigure}{\columnwidth}
        \centering
        \includegraphics{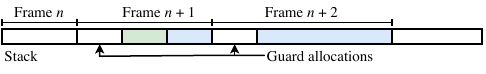}
        \caption{Stack layout with guard slots ensuring tag collisions never occur for adjacent allocations.}
        \label{fig:stack-layout}
    \end{subfigure}
    \caption{Untagged slots serving as guard slots to prevent tag collisions for adjacent allocations.}
    \label{fig:guard-allocs}
\end{figure}

\myparagraph{Heap safety}
\label{subsubsec:heap-safety}
In \projectname{}, the memory allocator must be aware of segments to provide heap safety.
Adapting a memory allocator to utilize \projectname{}'s memory safety extensions requires minimal modifications.
When allocating memory, the allocator must align the requested size to 16\,bytes, perform the allocation, and create a segment.
The corresponding tagged pointer is then returned to the caller. \projectname{} randomly selects a tag for each allocation. 
When freeing memory, \projectname{}'s allocator uses the \texttt{segment.free} instruction that ensures the detection of potential use-after-free and double-free errors.

Further, \projectname{} has to ensure that there are no tag collisions for adjacent allocations to provide protection against off-by-one buffer overflows/underflows.
This is achieved by design as \projectname{} places metadata at the beginning of every heap allocation, and the corresponding memory region is preserved untagged, as shown in \cref{fig:heap-layout}.
Thus, adjacent allocations are always separated by an untagged memory segment.

\myparagraph{Stack safety}
\label{subsubsec:stack-safety}
To provide memory safety for the stack, \projectname{} creates segments for the stack slots when entering a function.
\projectname{} generates a random tag per function for the first stack allocation. Subsequent stack allocations use this tag and increment it by one.
As the available tag bits are limited, the tag wraps around on overflow.
Before returning from a function, all stack slots are untagged and reassigned to the original stack frame.
This allows other functions to use the memory and prevents stack slots from being accessed after returning from a function.
Note that each stack allocation needs to be aligned to 16\, bytes and gets processed when entering/exiting from a function.

Similarly to the heap, \projectname{} must guarantee protection for off-by-one overflows/underflows on the stack.
Therefore, \projectname{} must ensure that two adjacent stack allocations between functions do not share the same tag.
To achieve this, \projectname{} inserts a single untagged stack guard slot at the beginning of the frame if no such untagged stack slot exists, as shown in \cref{fig:stack-layout}.
Without the guard allocation, adjacent allocations in stack frames $n+1$ and $n+2$ would share the same tag (blue) and, thus, \projectname{} would not be able to detect overflows.

To further reduce its performance and memory overheads, 
\projectname{} omits the instrumentation of stack allocations that \emph{(i)}~do not escape the function or \emph{(ii)}~are only accessed using statically verifiable indices. 
In \cref{alg:stack-safety-pseudo}, we present a simplified version of \projectname{}'s algorithm for the identification of the (un)safe stack allocations.

\LinesNumberedHidden
\setlength{\algomargin}{0pt}
\begin{algorithm}[t]
\myfontsizealgorithms
\SetAlgoLined
\SetKwInOut{KwIn}{Input}
\SetKwInOut{KwOut}{Output}
\KwIn{~Allocations}
\KwOut{~Hardened stack allocations}
\underline{{\bf handle\_stack\_allocations($\mathit{allocations}$)}} \\
\Begin{
    $\mathit{allocsToInstrument} \gets \emptyset$\\
    \ForEach{$\mathit{alloc} \in \mathit{allocations}$} {
        \uIf{escapes($alloc$)}
        {
            $\mathit{allocsToInstrument} \gets \mathit{allocsToInstrument} \cup \{\,\mathit{alloc}\,\}$
        }
        \ElseIf{isUsedByUnsafeGEP($alloc$)}
        {
            $\mathit{allocsToInstrument} \gets \mathit{allocsToInstrument} \cup \{\,\mathit{alloc}\,\}$
        }
    }

    \ForEach{$\mathit{alloc} \in \mathit{allocsToInstrument}$}{
        insertTaggingCode($\mathit{alloc}$)\\
        insertUntaggingCode($\mathit{alloc}$)\\
    }
    \If{$\mathit{allocsToInstrument} \neq \emptyset \land \mathit{allocations}[0] \notin \mathit{allocsToInstrument}$}{
        insertGuardAlloc()
}
}
\caption{Detect and harden safe and unsafe stack allocations.}

\label{alg:stack-safety-pseudo}

\end{algorithm}

\begin{figure}[t]
    \centering
    \includegraphics[width=\columnwidth]{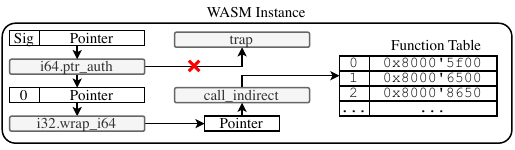}
    \caption{Our modified instruction sequence for indirect function calls.}
    \label{fig:ptr-auth-flow}
\end{figure}

\begin{figure}[t]
    \begin{equation*}
        \frac{C_{\text{memory}} = n}{C \vdash \textbf{segment.new}\ o : \text{i64}\ \text{i64} \rightarrow \text{i64}}
    \end{equation*}
    \begin{equation*}
        \frac{C_{\text{memory}} = n}{C \vdash \textbf{segment.set\_tag}\ o : \text{i64}\ \text{i64}\ \text{i64} \rightarrow \epsilon}
    \end{equation*}
    \begin{equation*}
        \frac{C_{\text{memory}} = n}{C \vdash \textbf{segment.free}\ o : \text{i64}\ \text{i64} \rightarrow \epsilon}
    \end{equation*}
    \begin{equation*}
        \frac{}{C \vdash \textbf{i64.pointer\_sign} : \text{i64} \rightarrow \text{i64}}
    \end{equation*}
    \begin{equation*}
        \frac{}{C \vdash \textbf{i64.pointer\_auth} : \text{i64} \rightarrow \text{i64}}
    \end{equation*}
    \caption{Typing rules of \projectname{}'s new instructions. For the definition of context $C$, see the \ac{WASM} paper~\cite{haas2017bringing}.}
    \label{fig:typing-rules}
\end{figure}

\myparagraph{Pointer authentication}
\projectname{} provides pointer authentication primitives that prevent function pointer reuse between \ac{WASM} instances.
On the instantiation of a \ac{WASM} module, a secret key is generated. The key is not accessible by the user code.
\projectname{}'s authentication operations leverage this key to sign and authenticate pointers using a cryptographic hash function.
The signature is placed in the unused 16\,bits of a \ac{WASM} pointer, alongside the pointer tag, if applicable.
Pointers containing a signature cannot directly access memory.
On authentication, the signature is checked and stripped, if it is valid. Otherwise, the \ac{WASM} module traps.
While function pointer reuse within a \ac{WASM} instance is still possible, \projectname{} prevents the reuse across different instances, as each instance generates its own key. \\
While memory64~\cite{memory64} extends function pointers to 64\,bits, the indices for the \ac{WASM} function table remain 32\,bit wide.
\projectname uses the instruction sequence in \cref{fig:ptr-auth-flow} to authenticate function pointers and perform indirect function calls.
64-bit pointers are first authenticated, which traps in case of a invalid signature.
If successful, the signature is stripped and the pointer is truncated to 32\,bits.
Similarly, when creating function pointers, indices into the function table are first zero-extended to 64\,bits and then signed.

\section{Semantics}

\begin{figure*}
    \footnotesize
    \begin{minipage}{\textwidth}
        \begin{align*}
            \text{(store)}\ s &\Coloneqq \{\dots, \text{inst}\ \mathit{inst}^* \} \\
            \mathit{inst} &\Coloneqq \{\dots, \text{tag}\ \mathit{taginst}^*, \text{key}\ k_s\} \\
            \mathit{taginst} &\Coloneqq b^*
        \end{align*}
        \begin{align}
            s;(\textbf{i64.const}\ k)~(t\textbf{.load}\ a\ o)\ &\hookrightarrow_i\ (t\textbf{.const}\ b^*)\label{eq:smallstep-1ok}
            && \text{if}\ s_\text{tag}(i, k + o,\lvert t \rvert) = \text{tag}(k) \land s_\text{mem}(i,k+o, \lvert t \rvert) = b^*\\
            s;(\textbf{i64.const}\ k)~(t\textbf{.load}\ a\ o)\ &\hookrightarrow_i\ \textbf{trap} \label{eq:smallstep-1trap}
            && \text{otherwise} \\
            s;(\textbf{i64.const}\ k)~(t\textbf{.const}\ c)~(t\textbf{.store}\ a\ o)\ &\hookrightarrow_i\ s'; \epsilon \label{eq:smallstep-2ok}
            && \text{if}\ s_\text{tag}(i, k + o,\lvert t \rvert) = \text{tag}(k)\\
            &&& \quad \land\ s' = s\ \text{with}\ \text{mem}(i,k+o, \lvert t \rvert) = \text{bits}^{\lvert t \rvert}_t(c) \nonumber\\
            s;(\textbf{i64.const}\ k)~(t\textbf{.const}\ c)~(t\textbf{.store}\ a\ o)\ &\hookrightarrow_i\ \textbf{trap} \label{eq:smallstep-2trap}
            && \text{otherwise} \\
            s;(\textbf{i64.const}\ k)~(\textbf{i64.const}\ l)~(\textbf{segment.new}\ o)\ &\hookrightarrow_i\ s';(\textbf{i64.const}\ t) \label{eq:smallstep-3}
            && \text{if}\ t = \text{new\_tag}(k + o) \land s' = s\ \text{with}\ \text{tag}(i, k + o, l) = t \\
            s;(\textbf{i64.const}\ k)~(\textbf{i64.const}\ l)~(\textbf{segment.new}\ o)\ &\hookrightarrow_i\ \textbf{trap} \label{eq:smallstep-6}
            && \text{otherwise} \\
            s;(\textbf{i64.const}\ k)~(\textbf{i64.const}\ t)~(\textbf{i64.const}\ l)~(\textbf{segment.set\_tag}\ o)\ &\hookrightarrow_i\ s';\epsilon \label{eq:smallstep-4}
            && \text{if}\ s' = s\ \text{with}\ \text{tag}(i, k + o, l) = t \\
            s;(\textbf{i64.const}\ k)~(\textbf{i64.const}\ t)~(\textbf{i64.const}\ l)~(\textbf{segment.set\_tag}\ o)\ &\hookrightarrow_i\ \textbf{trap} \label{eq:smallstep-7}
            && \text{otherwise} \\
            s;(\textbf{i64.const}\ k)~(\textbf{i64.const}\ l)~(\textbf{segment.free}\ o)\ &\hookrightarrow_i\ s';\epsilon \label{eq:smallstep-free-ok}
            && \text{if}\ t = \text{free\_tag}(k + o) \land s_\text{tag}(i, k + o,\lvert t \rvert) = \text{tag}(k) \\
            &&&\quad \land \ s' = s\ \text{with}\ \text{tag}(i, k + o, l) = t \nonumber\\
            s;(\textbf{i64.const}\ k)~(\textbf{i64.const}\ l)~(\textbf{segment.free}\ o)\ &\hookrightarrow_i\ \textbf{trap} \label{eq:smallstep-free-trap}
            && \text{otherwise} \\
            s;(\textbf{i64.const}\ k)~\textbf{i64.pointer\_sign}\ &\hookrightarrow_i\ (\textbf{i64.const}\ k') \label{eq:smallstep-9} && \text{if}\ k' = \text{sign}(k, k_s) \\
            s;(\textbf{i64.const}\ k)~\textbf{i64.pointer\_auth}\ &\hookrightarrow_i\ (\textbf{i64.const}\ k') \label{eq:smallstep-10} && \text{if}\ k' = \text{strip}(k) \land k = \text{sign}(k', k_s) \\
            s;(\textbf{i64.const}\ k)~\textbf{i64.pointer\_auth}\ &\hookrightarrow_i\ \textbf{trap} \label{eq:smallstep-11} && \text{otherwise} %
        \end{align}
    \end{minipage}
    \caption{Small-step reduction rules of the new instructions and added rules for load/stores. See the \ac{WASM} paper~\cite{haas2017bringing} for the definitions of all rules and auxiliary constructs.}
    \label{fig:smallstep-rules}
\end{figure*}

\subsection{Typing Rules}
\label{subsec:typing}
\projectname{} extends the typing rules of the original \ac{WASM} paper~\cite{haas2017bringing}, as shown in \Cref{fig:typing-rules}. We adopt the notation used in previous work~\cite{pierce2002types}.
Specifically, the rules are of the form of $C \vdash e : \mathit{tf}$.
An instruction $e$ is valid under the context $C$, with $C_\text{memory}$ being used to access a context component, such as the memory.
The rule $C_\text{memory} = n$ ensures that the instruction can only be used when memory is declared.
The type $\mathit{tf} = t_1^* \rightarrow t_2^*$ describes how the instruction manipulates the operand stack.
The instruction $e$ expects an operand stack where it pops $t_1^*$ and pushes $t_2^*$.

\subsection{Small-Step Reduction Rules}
\label{subsec:small-step-reduction-rules}
\Cref{fig:smallstep-rules} highlights how \projectname{} extends the \ac{WASM} small-step reduction rules~\cite{haas2017bringing} using the notation established by previous work~\cite{plotkin1981structural}.
The lower part of \cref{fig:smallstep-rules} presents the new tag-aware load/store rules that replace the load/store rules from the \ac{WASM} paper as well as new rules tailored based on the introduced instructions.
To signal a trap, we reuse operators from the original \ac{WASM} rules, including the \texttt{trap} operator.
The store, $s$, is augmented with a storage mechanism that assigns a tag, $t$, to each 16-byte memory granule and a per-instance secret key $k_s$.
The key is unique per \ac{WASM} instance and ensures that leaked signatures cannot be used in another instance or in another run of the same instance.
Precisely, we use the following notation:

\begin{itemize}[noitemsep, leftmargin=*]
    \item $t = s_{\text{tag}}(i, \mathit{addr}, \mathit{len})$: Extracts the tag $t$ for a memory region in instance $i$ accessed at address $\mathit{addr}$ with length $\mathit{len}$, if the tag is the same for all bytes in the range $\left[\mathit{addr}, \mathit{addr} + \mathit{len}\right)$.
    \item $s' = s\ \text{with}\ \text{tag}(i, \mathit{addr}, \mathit{len}) = t$: Updates the state with new tags for the memory region at address $\mathit{addr}$ with length $\mathit{len}$, if $\mathit{addr}$ is aligned to 16\,bytes and the memory region is in bounds of the memory.
    \item $t = \text{tag}(\mathit{pointer})$: Extracts the tag from a tagged pointer.
    \item $t' = \text{new\_tag}(t)$: Creates a tagged pointer $t'$ from an untagged pointer $t$ to be used for a new segment.
    The tag is randomly chosen from a pool of tags.
    \item $t' = \text{free\_tag}(t)$: Creates a tagged pointer $t'$ for the purpose of freeing a segment.
    The tag is different from the tag stored in $t$.
    \item $k'=\text{sign}(k, k_s)$: Creates a cryptographic signature based on the index $k$ and a per-instance secret key $k_s$ and inserts it into the upper bits of $k$.
    \item $k'=\text{strip}(k)$: Removes the cryptographic signature from the upper bits of $k$.
\end{itemize}

\noindent
Further, \Cref{fig:smallstep-rules} presents the added and modified components in the rules.
Each reduction rule is depicted with the top of the operand stack and the state, $s$, on the left side, representing the pre-execution state. The resulting stack and state after the execution of the instruction is placed on the right side.
For rules with $\hookrightarrow_i$, $i$ represents the instance in which the instruction is executed.
\projectname{} modifies the semantics of load/store instructions by adding new rules for loads and stores in \cref{eq:smallstep-1ok,eq:smallstep-1trap,eq:smallstep-2ok,eq:smallstep-2trap}, which trap on tag mismatches.

\begin{figure*}[t]
    \centering
    \begin{subfigure}[t]{\columnwidth}
        \vskip 0pt
        \includegraphics[scale=0.95]{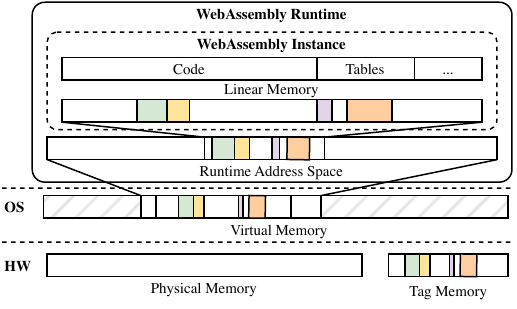}
        \caption{Internal memory safety, as implemented in wasmtime using \ac{MTE}. Memory segments are tagged, with different colors representing different tags. The virtual memory maps to physical and tag memory, which stores the tags assigned to memory granules.}
        \label{fig:wasmtime-mte-impl}
    \end{subfigure}
    \hfill
    \begin{subfigure}[t]{\columnwidth}
        \vskip 0pt
        \includegraphics[scale=0.95]{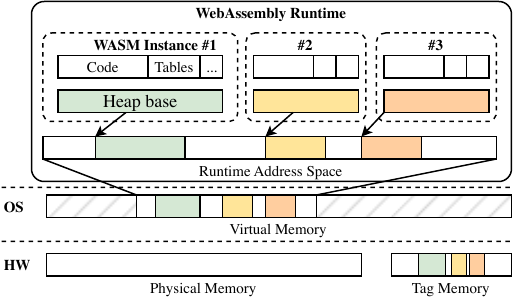}
        \caption{External memory safety enforced by \ac{MTE}. The linear memory of each instance is assigned a unique tag, which is stored in the heap base pointer and is used for effective address calculation.}
        \label{fig:system-design-sandboxing}
    \end{subfigure}
    \caption{System design of internal and external memory safety as implemented using \ac{MTE} in \projectname.}
    \label{fig:system-design-memory-safety}
\end{figure*}

Additionally, \projectname{} introduces new rules that express its new instructions, which are described below:
\begin{itemize}[noitemsep, leftmargin=*]
    \item \texttt{segment.new}: It creates a new, zeroed memory segment for a given pointer and size and returns a tagged pointer (\cref{eq:smallstep-3}).
    \item \texttt{segment.set\_tag}: It transfers ownership for a given point\-er and size to a tagged pointer (\cref{eq:smallstep-4}).
    This results in the tagged pointer being able to access the memory segment and can be used to merge adjacent segments.
    \item \texttt{segment.free}: It invalidates a segment specified by a pointer and length by tagging the segment with a new, implementation-defined tag (\cref{eq:smallstep-free-ok,eq:smallstep-free-trap}), ensuring accesses to the freed segment are caught.
    This instruction also traps if a segment is freed twice, i.e., the given tagged pointer cannot access the memory region.
    \item \texttt{i64.pointer\_sign}: It signs a pointer and places a cryptographic hash in its upper bits (\cref{eq:smallstep-9}).
    \item \texttt{i64.pointer\_auth}: It validates a signed pointer to ensure the hash in the upper bits matches the address in the lower bits.
    If the hash matches, the hash is removed from the upper bits (\cref{eq:smallstep-10}).
    Otherwise, it produces a trap (\cref{eq:smallstep-11}).
\end{itemize}

\noindent
Using the above rules, the \cref{eq:smallstep-6,eq:smallstep-7,eq:smallstep-free-trap} of \Cref{fig:smallstep-rules} ensure traps when code tries to create, modify, or free unaligned segments or segments outside the linear memory for a given instance.
\projectname{} requires all segments to be aligned to 16\,bytes.

\section{Implementation}
\label{sec:implementation}

\projectname{} is built based on LLVM 17, wasi-libc, and wasmtime 16.0.0.
\projectname consists of a modified LLVM toolchain~($\S$\ref{subsec:llvm}), wasi-libc~($\S$\ref{subsec:wasi-libc}) and implementation of the memory safety extension in wasmtime~($\S$\ref{subsec:internal-memory-safety-impl}, $\S$\ref{subsec:external-memory-safety-impl}).

\subsection{LLVM Compiler Toolchain}
\label{subsec:llvm}

In \projectname{}, we use LLVM~\cite{lattner2004llvm} to compile our C/C++ applications to WebAssembly.
We extend its existing \ac{WASM} backend to be able to emit the new \projectname{}'s instructions.
We further introduce new intrinsic functions that correspond and are lowered to \projectname{}'s \ac{WASM} instructions by the compiler.
The clang frontend and the \projectname{} compiler passes are mainly responsible for inserting calls to \projectname{}'s new intrinsic functions when necessary.
Precisely, we introduce two \ac{WASM}-specific sanitizer passes that can be enabled in the LLVM pipeline via compiler flags.
In clang, we introduce new built-in functions, which directly map to these intrinsics and allow programmers to use the new instructions, e.g., to build a segment-aware memory allocator.

The first sanitizer pass is designed to provide memory safety for stack allocations when compiling the source code to WebAssembly.
It analyzes functions for stack allocations, applies padding, and creates the memory segments ($\S$\ref{subsubsec:stack-safety}), ensuring temporal and spatial safety of stack allocations.

The second sanitizer pass enforces pointer authentication for indirect function calls.
\projectname instruments code taking references to functions and performing indirect calls.
We do not handle other operations on code pointers, as pointer arithmetic on function pointers is undefined behavior in C~\cite[\S6.5.6/2, \S6.5.2/1]{iso9899} and C++~\cite[\S7.6.6/1, \S6.8/8, \S31.8.4/4]{iso14882}.

Note that both sanitizer passes run after all LLVM optimizations.
This ensures that \projectname{} does not block passes that might remove stack allocations, such as \texttt{mem2reg}.

\begin{figure*}[t]
    \centering
    \begin{subfigure}{\columnwidth}
        \centering
        \includegraphics{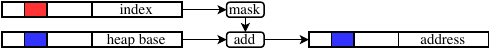}
        \caption{Sandboxing with \ac{MTE}, with all tags being reserved for the runtime to isolate \ac{WASM} guests.}
        \label{fig:ptr-masking-external}
    \end{subfigure}
    \hfill
    \begin{subfigure}{\columnwidth}
        \centering
        \includegraphics{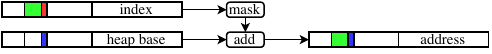}
        \caption{\Ac{MTE} sandboxing and the memory safety extension combined, with three tags reserved for the guest.}
        \label{fig:ptr-masking-combined}
    \end{subfigure}
    \caption{Pointer masking to ensure bounds tags cannot be manipulated.}
    \label{fig:ptr-masking}
\end{figure*}

\subsection{WASI Libc Modifications}
\label{subsec:wasi-libc}

We port the \acf{WASI} and wasi-libc to wasm64 to run applications relying on libc on wasm64. To achieve this, \projectname{} adapts the size and the pointer types from 32 to 64\,bits.
Further, we modify \texttt{dlmalloc}, the default allocator in wasi-libc, to provide memory safety for heap allocations.
\projectname{}' allocator creates memory segments and returns tagged pointers to these segments.
Thus, it protects both allocator metadata and adjacent allocations from being accessed or modified through heap overflows, as illustrated in \cref{fig:heap-layout}.
When freeing or reallocating memory, segments are freed, ensuring temporal safety. 
Lastly, we recompile WASI-libc with pointer authentication to ensure function pointers are signed and authenticated by library code.

\subsection{Internal Memory Safety}
\label{subsec:internal-memory-safety-impl}

\Cref{fig:wasmtime-mte-impl} illustrates an overview of \projectname{}'s implementation of the internal memory safety ($\S$\ref{subsec:system-model}) extension in wasmtime using \ac{MTE}.
We modify wasmtime and its supporting libraries to parse, process, and enforce the memory safety extension described in $\S$\ref{subsec:wasm-extension}.
More specifically, we add support for \ac{MTE} and \ac{PAC} in the form of new instructions and lowering rules.
We implement segments using \ac{MTE}.
\projectname{} provides memory safety for the heap via its modified allocator that protects heap allocations by creating segments.
Respectively, to provide memory safety for the stack, \projectname{} adapts the stack allocations; the first segment in each function is assigned a random tag and
successive allocations increment the tag by one ($\S$\ref{subsec:wasm-extension}), eventually wrapping around.
This ensures adjacent allocations on the stack never share a tag.
Importantly, \projectname{} uses synchronous \ac{MTE} to trap on memory safety violations before their effects become observable.

For pointer authentication, \projectname{} ensures that each \ac{WASM} instance receives a distinct secret key to sign pointers.
If multiple \ac{WASM} instances run in a single process, \projectname{} initializes a global random value per instance used as the PAC's instructions modifier.
This is required, as PAC keys are shared per process.
Otherwise, \projectname uses \ac{PAC} instructions without a modifier.

\subsection{External Memory Safety}
\label{subsec:external-memory-safety-impl}

\cref{fig:system-design-sandboxing} presents how \projectname{} utilizes memory tagging to replace software-based bounds checks and preserve external memory safety ($\S$\ref{subsec:system-model}) for each \ac{WASM} instance.
Initially, the runtime assigns a tag to each instance on module instantiation, which is stored in the heap base address.
The effective memory access address is calculated by adding the accessed index to the tagged heap base address.
The memory of the runtime is tagged with zero. 
It enables \ac{MTE} to catch any access outside the sandbox due to the tag mismatch. 

\projectname{} can further combine both its \emph{external} and its \emph{internal} memory safety extensions by splitting the pointer tag bits among them.
The upper bits are used for internal memory safety, while the lower bits are reserved for sandboxing.
In \projectname{}'s prototype, we assign three bits for internal memory safety and one bit for enforcing external memory safety.
This isolates a single \ac{WASM} instance while assigning three bits for internal memory safety.

Lastly, \projectname{} must ensure that adding an untrusted pointer to the heap base does not allow \ac{WASM} code to craft arbitrary values to escape the sandbox.
To provide these guarantees, \projectname{} masks the \ac{WASM} index before address computation, as shown in \cref{fig:ptr-masking}.
\projectname masks bits 56--59 when only external memory safety is enabled (\cref{fig:ptr-masking-external}) and bit 56 when both internal and external memory safety are enabled (\cref{fig:ptr-masking-combined}).

\myparagraph{Limitations}
A natural limitation of \projectname{} is the number of sandboxes within a single process. \ac{MTE} provides a limited number of tags (16). \projectname{} reserves the zero tag for the runtime, leaving the remaining 15 for the sandboxes. 
In a future version of \projectname{}, the number of available sandboxes can be increased by ensuring tags are only reused for memory regions outside the address range reachable by \ac{WASM} pointers, i.e., via combining guard pages with memory tagging.

We must exclude certain tags from being generated by \ac{MTE} instructions, as \projectname{} uses the tag bit 56 to distinguish between runtime and guest memory.
On top of that, \projectname{} must also exclude the zero tag from the guest, as this tag is reserved for guard slots and untagged segments.
To this end, at runtime startup, we specify which tags can be generated using the \texttt{prctl} mechanism.

\begin{table}[t]
    \centering
    \smaller
    \caption{Runtime benchmarking configurations.}
    \label{tab:benchmark-variants}
    \begin{tabular}{l || l|l|l|l}
        \textbf{Variant} & \textbf{Ptr width} & \textbf{Internal} & \textbf{External} & \textbf{Ptr auth} \\
        \hline
        baseline wasm32           & 32-bit           & No                 & No                & No              \\
        baseline wasm64           & 64-bit           & No                 & No                & No              \\
        \projectname{}-mem-safety & 64-bit           & Yes                & No                & No              \\
        \projectname{}-ptr-auth   & 64-bit           & No                 & No                & Yes             \\
        \projectname{}-sandboxing & 64-bit           & No                 & Yes               & No              \\
        \projectname{}            & 64-bit           & Yes                & Yes               & Yes             \\
    \end{tabular}
\end{table}

\section{Evaluation}
\label{sec:eval}

\subsection{Experimental Setup}\label{subsec:experimental-setup}

We conduct our benchmarks on a Google Pixel 8 equipped with an ArmV9 Tensor G3 chip, including one Cortex-X3 (2.91\,GHz, out-of-order), four Cortex-A715 (2.37\,GHz, out-of-order), and four Cortex-A510 (1.7\,GHz, in-order) cores.
All cores feature \ac{MTE} and \ac{PAC} with FEAT\_FPAC enabled, making \ac{PAC} trap on authentication.

\myparagraph{Methodology}
We evaluate \projectname on the PolyBench/C 3.2 suite~\cite{polybenchc} and microbenchmarks with the configurations outlined in \cref{tab:benchmark-variants}.
These include wasm32 and wasm64 as baselines and the memory-safety, pointer-authentication ($\S$\ref{subsec:internal-memory-safety-impl}), and sandboxing ($\S$\ref{subsec:external-memory-safety-impl}) components from \projectname{}, as well as all components of \projectname{} combined.
We perform and compare each benchmark on every type of CPU core available on the Tensor G3 by pinning benchmarks to a single core.

\begin{figure}[t]
    \centering
    \includegraphics[width=1.0\columnwidth]{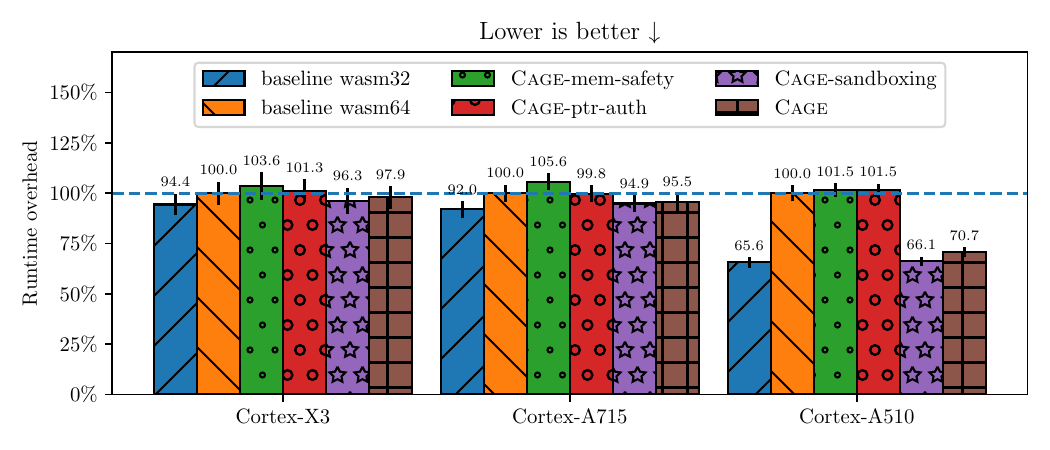}
    \caption{PolyBench/C runtime overheads of different configurations described in \cref{tab:benchmark-variants}, normalized to wasm64.}
    \label{fig:runtime-overheads-combined}
\end{figure}

\subsection{Performance Overheads}
\label{subsec:performance-overheads}

\Cref{fig:runtime-overheads-combined} illustrates the mean runtime overheads for the PolyBench/C benchmarks. \\
\myparagraph{\projectname memory safety} Compared to wasm64, our memory safety extension has a mean overhead of $3.6 \pm 6.9\,\%$, $5.6 \pm 4.3\,\%$, and $1.5 \pm 3.3\,\%$.
\myparagraph{\projectname sandboxing} \Ac{MTE}-based sandboxing achieves a mean speedup of $3.7 \pm 6.5\,\%$, $5.1 \pm 4.0\,\%$, and $33.9 \pm 2.4\,\%$ over wasm64.
\myparagraph{\projectname} When combining our memory safety extension with \ac{MTE}-based sandboxing, we see a mean speedup of $2.1 \pm 5.6\,\%$, $4.5 \pm 4.1\,\%$, and $29.2 \pm 2.5\,\%$ compared to wasm64, while providing stronger security guarantees.

\begin{figure}[t]
    \centering
    \includegraphics[width=1\columnwidth]{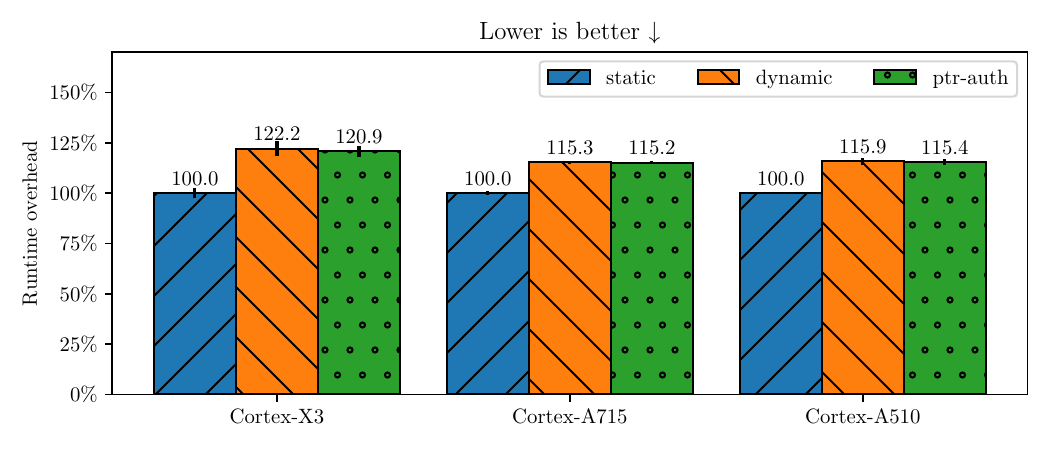}
    \caption{Overheads of pointer authentication.}
    \label{fig:runtime-pauth}
\end{figure}

\myparagraph{\projectname pointer authentication} As the PolyBench/C test suite does not exercise virtual function calls, we only see an overhead within error margins over 64-bit \ac{WASM}.
We perform a microbenchmark, comparing static against dynamic and authenticated dynamic function calls in \cref{fig:runtime-pauth}.
We observe virtually no overhead occurring as the result of pointer authentication, as adding pointer authentication only adds 5 cycles of latency, which is not noticeable.
Switching from static to dynamic function calls results in a 15\,\%--22\,\% overhead, depending on the CPU being run on.

\myparagraph{\projectname Startup Overhead}
We measure the startup overhead for a \ac{WASM} instance with a static memory of 128\,MiB, calling a function that immediately returns.
The overhead of tagging the linear memory is hidden by the runtimes startup overhead.

\subsection{Memory Overheads}\label{subsec:memory-overheads}

The memory overheads of \projectname consist of two factors:
\emph{(i)} the overhead of switching from 32-bit to 64-bit \ac{WASM} and
\emph{(ii)} the overhead of memory tags caused by \ac{MTE}, which is $1/32$ of the memory with \ac{MTE} enabled.
Tags are stored in a separate physical address space, the \textit{tag PA space}~\cite{ARMA2024Arch64}.
This space is not managed by the operating system and is not included in maximum \ac{rss} measurements.
We thus estimate the memory overhead with \ac{MTE} by factoring in 4 bits for every 16\,byte of memory, which results in an additional $1/32=3.125\,\%$.
As our maximum \ac{rss} measurements include the runtimes memory, which does not use \ac{MTE}, this should be considered an over-approximation; in reality, the memory overhead of \projectname is lower.
We measure the mean overhead of wasm64 over wasm32 at $0.6\,\%$ and thus estimate \projectname's overhead at $<5.3\,\%$.

\subsection{Security Guarantees}\label{subsec:security-guarantees}

We evaluate the security guarantees from the perspective of internal and external memory safety (\cref{par:threat-model}).

\myparagraph{External memory safety}
\label{subsubsec:sec-guarantees-external-memory-safety}
\projectname implements sandboxing with \ac{MTE} in synchronous mode, which enforces sandboxes through hardware.
This prevents sandbox escapes, such as CVE2023-26489, from accessing memory outside the linear memory.
We prevent malicious code from forging tags to escape the sandbox, as described in $\S$\ref{subsec:external-memory-safety-impl}.
However, we limit the number of sandboxes in one process to at most 15, which is required to assign a distinct tag to each sandbox and one to the runtime.

\myparagraph{Internal memory safety}
\label{subsubsec:sec-guarantees-internal-memory-safety}
Our choice to utilize a memory-tagging-based approach does not provide complete memory safety, as we rely on a limited number of tags.
As we reserve one tag for guard allocations and untagged segments, the probability for a tag collision is $1/15=6.\overline{6}\,\%$.
If we utilize \ac{MTE} for sandboxing, the chance of a tag collision rises to $1/7 \approx 14.3\,\%$.
\projectname deterministically protects against off-by-one overflows, use-after-free, and double-free errors, which are caught at least until the reuse of a memory allocation. \\
Signing function pointers further reduces the possible attack surface.
As \ac{WASM} already provides strong guarantees against control-flow-based attacks, adding pointer authentication primitives does not substantially improve security.
\projectname protects against reusing leaked function pointers between instances and statically deducing function pointers. \\
We recommend \projectname to be used as a secondary defense mechanism to mitigate classes described in \cref{tab:cves}.
With the ability to deploy the prototype in production, bugs may be found in production environments not discovered by testing workloads.

\begin{table}[t]
    \centering
    \small
    \caption{Variants for initializing and tagging memory.}
    \label{tab:stg-instructions}
    \begin{tabular}{c || c | c | c | c }
        \textbf{Variant} & \textbf{Instr} & \textbf{Granule} & \textbf{Sets 0} & \textbf{memset} \\
        \hline
        memset           & -              & -                     & No              & Yes             \\
        stg              & \texttt{stg}   & 16\,bytes             & No              & No              \\
        st2g             & \texttt{st2g}  & 32\,bytes             & No              & No              \\
        stgp             & \texttt{stgp}  & 16\,bytes             & Yes             & No              \\
        stzg             & \texttt{stzg}  & 16\,bytes             & Yes             & No              \\
        st2zg            & \texttt{st2zg} & 32\,bytes             & Yes             & No              \\
        stg+memset       & \texttt{stg}   & 16\,bytes             & Yes             & Yes             \\
        st2g+memset      & \texttt{st2g}  & 32\,bytes             & Yes             & Yes             \\
    \end{tabular}
\end{table}

\begin{figure}[t]
    \centering
    \includegraphics[width=1.0\columnwidth]{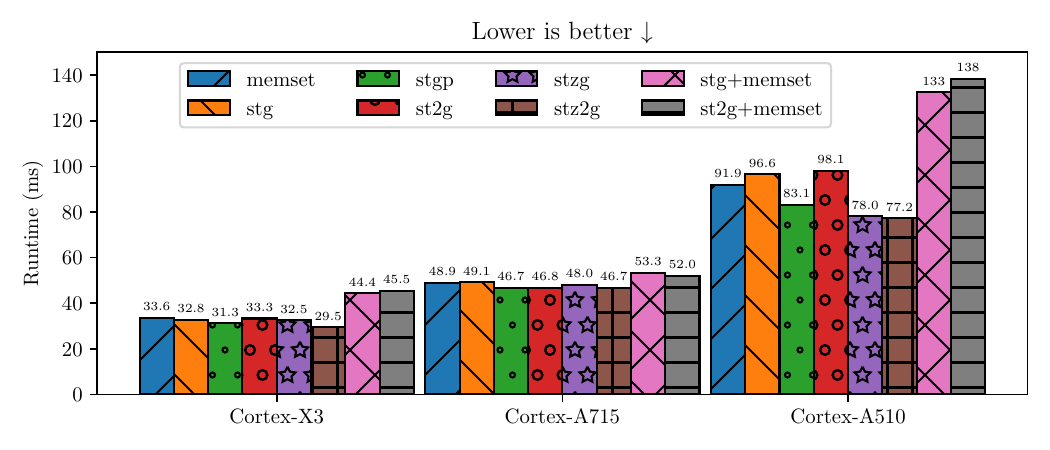}
    \caption{Performance results of the benchmarking variants from \cref{tab:stg-instructions} on 128\,MiB of memory.}
    \label{fig:stg-performance}
\end{figure}

\myparagraph{Initializing tagged memory}
\label{subsubsec:tagging-primitives}
We benchmark several primitives to initialize an uncached 128\,MiB large memory region while setting its allocation tag, representing the scenario of setting up a \ac{WASM} instance.
We run the configurations in \cref{tab:stg-instructions} with synchronous \ac{MTE} and measure the results in \cref{fig:stg-performance}.
\texttt{stzg}, \texttt{stz2g}, and \texttt{stgp} are slightly faster than a raw memset, even though they initialize memory and set tags.
We believe this is the case as they do not perform tag checks before accessing memory~\cite{ARMA2024Arch64}.
The Cortex-X3 core compensates for its lower throughput (\cref{tab:instruction-latencies}) with a higher clock speed (2.91\,GHz) compared to the Cortex-A715 (2.37\,GHz).

\section{Related Work}
Prior research exists on enriching \ac{WASM} with memory safety properties.
A significant project is \emph{MS-WASM}~\cite{disselkoen2019position,michael2023mswasm}, a memory safety extension for \ac{WASM} that introduces a new \textit{segment memory} distinct from the linear memory, preventing access through arbitrary offsets.
The segment memory relies on accesses through unforgeable \textit{handles}, akin to CHERI pointers~\cite{woodruff2014cheri}.
On the contrary, \projectname{} does not introduce a distinct memory region and a new pointer type, but rather builds upon 64-bit \ac{WASM} without altering its memory layout to be able to minimize its runtime overheads on devices supporting Arm's \ac{MTE}.

\emph{RichWasm} is another approach towards memory safety for \ac{WASM}~\cite{paraskevopoulou2024richwasm}.
It provides a richly typed intermediate language for safe memory interactions between languages with varying memory management models.
RichWasm allows for static detection of memory safety violations in the context of mixed-language interoperability between strongly typed languages like Rust or OCaml. %
Unsafe languages (e.g., C/C++) lack information for static safety analysis and are not directly supported by RichWasm's type-driven safety model.
In contrast, \projectname{} targets languages that do not encode safety in their runtime or type system.

\emph{Wasm memcheck (wmemcheck)}~\cite{wmemcheck} is a tool in wasmtime providing the ability to check for invalid mallocs, double-frees, reads, and writes inside a \ac{WASM} module, assuming certain properties about \texttt{malloc} and \texttt{free}.
It is conceptually similar to Valgrind memcheck~\cite{nethercote2007valgrind}.

\balance
\section{Conclusion}
\label{sec:conclusion}

In this paper, we present \projectname, a hardware-accelerated \acl{WASM} toolchain to provide memory safety, consisting of a minimally invasive and adaptable \ac{WASM} extension, a compiler toolchain based on LLVM, a modified wasi-libc that includes a custom allocator to provide spatial and temporal memory safety, and a wasmtime implementation, that is responsible for compiling and running the \projectname{}'s \ac{WASM} extension using \ac{MTE} and \ac{PAC}.
Further, \projectname{} improves the performance of \ac{WASM} sandboxing mechanism by utilizing \ac{MTE} as a replacement for software-based bounds checks.
Our evaluation of \projectname{} highlights that \projectname{} is a memory safety solution for \ac{WASM} that is suitable for production deployment as it incurs minimal runtime and memory overheads while providing efficient memory safety guarantees.

\myparagraph{Software artifact} \projectname{} is publicly available with its entire experimental setup~\cite{artifact}.
Detailed information can be found in \cref{subsec:ae-description}.

\section*{Acknowledgement}
We thank the anonymous reviewers for their helpful feedback.
We also thank Fritz Rehde, Janne Mantyla, and Carlos Chinea Perez for their work and feedback in the early stages of the project.
This work was partially supported by a research grant from Huawei Research Finland, the TUM Innovation Network Resilient, Trustworthy, Sustainable (ReTruSt), a Google Safe Compilation grant, and an ERC Starting Grant (ID: 101077577).

\newpage

\appendix
\section{Artifact Appendix}

\subsection{Abstract}

This appendix provides the necessary information to build the artifacts and reproduce the experiments of the CGO'25 paper ``\projectname: Hardware-Accelerated Safe WebAssembly'' by M. Fink, D. Stavrakakis, D. Sprokholt, S. Chakraborty, J.-E. Ekberg., and P. Bhatotia.
\projectname provides a memory safety abstraction for WebAssembly, with an implementation in LLVM to transparently compile unmodified C/C++ programs, a modified libc that provides memory safety for heap allocations, and an implementation in wasmtime that utilizes Arm MTE and PAC to implement the abstraction.
We provide an artifact including pre-compiled binaries and all sources and scripts to build and evaluate the artifact to reproduce the results and figures in this paper.

\subsection{Artifact Check-List (Meta-Information)}

{\small
\begin{itemize}
  \item \textbf{Program:} LLVM with \projectname modifications, source included; wasmtime with \projectname modifications, source included; wasi-libc with \projectname modifications, source included.
  \item \textbf{Compilation:} Clang 17, rustc 1.80
  \item \textbf{Transformations:} Stack allocation hardening as an LLVM pass.
  \item \textbf{Binary:} Pre-built binaries for wasmtime, LLVM, wasi-sdk included. Source code and makefiles to re-generate binaries included.
  \item \textbf{Run-time environment:} Provided binaries built for Linux 6.8.12 (nixOS) and Android 14.
  \item \textbf{Hardware:} We require an AArch64 device with both PAC and MTE (Pixel 8) and an x86-64 machine for cross-compilation.
  \item \textbf{Metrics:} Average runtime overhead; estimated memory overhead
  \item \textbf{Output:} PDFs for the plots; Text files with raw data for tables and claims in text.
  \item \textbf{Experiments:} Makefile to run all experiments is included.
  \item \textbf{How much disk space required (approximately)?:} 25 GiB.
  \item \textbf{How much time is needed to prepare workflow (approximately)?:} 2--3 hours.
  \item \textbf{How much time is needed to complete experiments (approximately)?:} 2--3 days.
  \item \textbf{Publicly available?:} Yes.
  \item \textbf{Code licenses (if publicly available)?:} Apache License with LLVM Exceptions (LLVM, wasi-libc); Apache License (wasmtime).
  \item \textbf{Workflow framework used?:} Makefiles.
  \item \textbf{Archived (provide DOI)?:} \href{https://doi.org/10.5281/zenodo.13772996}{10.5281/zenodo.13772996}
\end{itemize}

\subsection{Description}
\label{subsec:ae-description}

\subsubsection{How Delivered}

All source code can be found at the git repositories below, as well as the following persisted DOI for the artifact: \href{https://doi.org/10.5281/zenodo.13772996}{10.5281/zenodo.13772996}, which contains all source code as well as the scripts to build and evaluate all artifacts.
\begin{itemize}
  \item \url{https://github.com/TUM-DSE/llvm-memsafe-wasm}
  \item \url{https://github.com/TUM-DSE/wasmtime-mte}
  \item \url{https://github.com/TUM-DSE/wasm-tools-mte}
  \item \url{https://github.com/martin-fink/wasi-libc}
\end{itemize}

\subsubsection{Hardware Dependencies}

The evaluation is performed on a Google Pixel 8 with Arm \ac{MTE} and \ac{PAC}.
We cross-compile LLVM, wasmtime, and the benchmarks on an x86 machine.

\subsubsection{Software Dependencies}

We require the following set of software on the Google Pixel 8 device:
\begin{itemize}
  \item Termux
  \item sshd
  \item bash
\end{itemize}

We require the following set of software on the x86 machine to compile LLVM, wasmtime, and evaluate the benchmarks:

\begin{itemize}
  \item Linux (tested with 6.8.12)
  \item nix (tested with 2.18.5): All other dependencies are fetched and pinned to a specific version using the nix package manager and can be found in the \texttt{nix/} directory in the artifact.
\end{itemize}

\subsubsection{Benchmarks}

We run the PolyBench/C benchmark suite to measure the runtime and memory overhead of \projectname's components compared to the 32- and 64-bit baselines.
To measure startup overheads, we measure the overhead of instantiating a module declaring a 128\,MiB memory and calling an empty function.
To measure pointer authentication overheads, we measure a modified version of PolyBench/C's 2mm benchmark, where the matrix multiplication is moved into a function call that is either performed statically or dynamically through a vtable.

\subsection{Installation}

To get started, download the artifact from Zenodo, navigate to the artifact directory, and run the following command to download all required dependencies using nix.

\begin{lstlisting}[frame=h,style=custom]
curl -L -o cgo-artifact.zip https://zenodo.org/records/13772996/files/cgo-artifact.zip?download=1
nix-shell -p unzip --run 'unzip cgo-artifact.zip'
cd cgo-artifact
cd nix
nix develop
cd ..\end{lstlisting}

This opens a new shell with all dependencies required to build, run, and evaluate all benchmarks.

\subsubsection{SSH Connection to the Pixel 8}

Install Termux from the Play Store or F-Droid.
Once Termux is opened, install and start sshd, then connect to the x86 machine and open a port forwarding, allowing the x86 machine to connect to the Pixel 8.

\begin{lstlisting}[frame=h,style=custom]
pkg install sshd
sshd
ssh -R 8023:localhost:8022 user@x86machine\end{lstlisting}

To connect to the Pixel 8, replace the following two lines in \texttt{config.mk} with the values corresponding to your device.

\begin{lstlisting}[frame=h,style=custom]
export SSH_HOST=u0_a265@localhost
export SSH_PORT=8023\end{lstlisting}

\subsection{Experiment Workflow}

All required software (LLVM, wasmtime, benchmarks) is cross-compiled on the x86 machine and copied to the Pixel 8 using a set of provided scripts.

\paragraph{Building LLVM, wasmtime, wasi-sdk, and the Benchmarks:} \emph{[1 human-minute + 2--3 compute-hours]}

For convenience and to reduce build times, we have included pre-built versions of wasi-sdk and LLVM in the artifact.
To build them from scratch, delete the following files:

\begin{lstlisting}[frame=h,style=custom]
rm -rf toolchain/wasi-sdk-20
rm -rf toolchain/wasi-sdk-20+memory64
rm -rf toolchain/wasi-sdk-20+memory64+memsafety
rm -rf toolchain/wasi-sdk-20+memory64+memsafety+ptr-auth
rm -rf toolchain/wasi-sdk-20+memory64+ptr-auth\end{lstlisting}

To build the toolchain and benchmarks, run:

\begin{lstlisting}[frame=h,style=custom]
make -j$(nproc) build\end{lstlisting}

This produces the following artifacts, as well as the benchmarks used in this paper:

\begin{lstlisting}[frame=h,style=custom]
# wasmtime:
./toolchain/wasmtime/target/aarch64-linux-android/release/wasmtime
# llvm:
./toolchain/wasi-sdk-20+memory64+memsafety+ptr-auth/wasi-sdk-wasi-sdk-20+memory64+memsafety+ptr-auth/bin/clang
# wasi-sdk with different configurations
./toolchain/wasi-sdk-20
./toolchain/wasi-sdk-20+memory64
./toolchain/wasi-sdk-20+memory64+memsafety
./toolchain/wasi-sdk-20+memory64+ptr-auth
./toolchain/wasi-sdk-20+memory64+memsafety+ptr-auth
\end{lstlisting}

\subsection{Evaluation and Expected Result}

\textbf{Expected duration:} \emph{[1 human-minute + 2--3 compute-days]}

\noindent
Running the experiments on the Pixel 8 devices takes a long time.
This is primarily caused by our choice to run all experiments on all three types of cores found in the Pixels chipset.
Running the benchmarks on the low-power Cortex-A510 cores takes up most of the runtime.
To perform the evaluation, run:

\begin{lstlisting}[frame=h,style=custom]
make -j$(nproc) evaluate\end{lstlisting}

This copies all benchmarks and artifacts to the connected Pixel 8, performs benchmarks, copies the results to the x86 machine, and creates plots and results claimed in the paper.
The results can be found in the \texttt{results/} directory.
We reproduce the following figures/claims:
\begin{itemize}
  \item Runtime overhead (\cref{fig:runtime-overheads-combined}): \texttt{results/runtime.pdf}
  \item Pointer auth. overhead (\cref{fig:runtime-pauth}): \texttt{results/ptr-auth.pdf}
  \item Memory overhead (\cref{subsec:memory-overheads}): \texttt{results/mem.txt}
  \item Startup overhead (\cref{subsec:performance-overheads}): \texttt{results/startup.txt}
  \item Memory tagging overheads (\cref{fig:stg-performance}): \texttt{results/stg.pdf}
\end{itemize}

\noindent
Additionally, we reproduce the following architectural analysis results from $\S$\ref{sec:background}:
\begin{itemize}
  \item MTE sync/async mode overhead (\cref{fig:sync-async-performance}):\\ \texttt{results/mte-mode.pdf}
  \item MTE instruction latencies/throughput (\cref{tab:instruction-latencies}):\\ \texttt{results/inst-cycles.txt}
\end{itemize}

\subsection{Notes}

We provide hardware with the required software preinstalled for the CGO'25 artifact reviewers.

\subsection{Methodology}

Submission, reviewing and badging methodology:

\begin{itemize}
  \item \url{http://cTuning.org/ae/submission-20190109.html}
  \item \url{http://cTuning.org/ae/reviewing-20190109.html}
  \item \url{https://www.acm.org/publications/policies/artifact-review-badging}
\end{itemize}

\balance

\newpage
\balance
\bibliographystyle{ACM-Reference-Format}
\bibliography{main}

\end{document}